\pdfoutput=1
\documentclass[a4paper]{article}
\usepackage{amsmath}
\usepackage{amssymb}
\usepackage{authblk}
\usepackage[english]{babel}
\usepackage[utf8]{inputenc}
\usepackage[
    sortcites,
    backend=biber,
    giveninits=true,
    minbibnames=3,
    style=ext-numeric-comp,
    articlein=false,
    ]{biblatex}
\usepackage{csquotes}
\usepackage[T1]{fontenc}
\usepackage{microtype}
\usepackage{tikz}
\pgfsyspdfmark {pgfid27}{42}{420000}
\pgfsyspdfmark {pgfid30}{42}{42}
\pgfsyspdfmark {pgfid32}{42}{42}
\pgfsyspdfmark {pgfid37}{42}{420000}

\usepackage[disable]{todonotes} %
\usepackage{multirow}
\usepackage{booktabs}
\usepackage{caption}
\usepackage{url}
\usepackage{hyperref}

\usetikzlibrary{arrows.meta}
\usetikzlibrary{backgrounds}
\usetikzlibrary{fit}
\usetikzlibrary{shapes}
\usetikzlibrary{positioning}
\usetikzlibrary{quotes}
\usetikzlibrary{calc}

\tikzset{contour/.style={black!60, dashed, shorten <=-1.1cm, rounded corners=.5mm, thick}}
\tikzset{edge/.style={-{Stealth[scale=1.2]}, shorten >=1pt, semithick}}
\tikzset{lambda edge/.style={dashed, edge}}
\tikzset{vertex/.style={circle, draw, minimum size=1.6cm, thick}}
\tikzset{vertexsm/.style={circle, draw, minimum size=1cm, thick}}
\newcommand*\variant[3]{$#1\!:\!#2\,/\,\texttt{#3}$}  %

\definecolor{highlight}{HTML}{FF8C00}
\definecolor{orange}{HTML}{FF8C00}
\definecolor{green}{HTML}{006400}
\definecolor{blue}{HTML}{003366}
\definecolor{purple}{HTML}{FF00F3}

\newcommand{\tikzmark}[1]{\tikz[overlay, remember picture] \node (#1) {};}
\newcommand*\tikzcircled[2][char]{\tikz[baseline=(#1.base), overlay, remember picture]{\node[circle, draw, inner sep=.4pt] (#1) {$#2$};}}

\AtBeginBibliography{\small}
\author[1,2,*]{Mark~A.~Santcroos}
\author[3]{Walter~A.~Kosters}
\author[1]{Mihai~Lefter}
\author[1,4]{Jeroen~F.J.~Laros}
\author[1,3]{Jonathan~K.~Vis}
\affil[1]{Department of Human Genetics, Leiden University Medical Center,
The Netherlands}
\affil[2]{Department of Clinical Genetics, Leiden University Medical Center,
The Netherlands}
\affil[3]{Leiden Institute of Advanced Computer Science, Leiden University,
The Netherlands}
\affil[4]{National Institute for Public Health and the Environment (RIVM),
The Netherlands}

\affil[*]{\href{mailto:m.a.santcroos@lumc.nl}{m.a.santcroos@lumc.nl}}

\title{A Graph-based Approach to Variant Description Extraction from Sequences}

\begin{document}
\maketitle

\begin{abstract}
\noindent
Accurate variant descriptions are of paramount importance in the field of
genomics.
The domain is confronted with increasingly complex variants, e.g.,
combinations of multiple indels, making it challenging to generate proper
variant descriptions directly from chromosomal sequences.

\noindent
We present a graph based on all minimal alignments that is a complete
representation of a variant, which gives insight into the nature of
a variant compared to a single variant description.
We provide three complementary extraction methods to derive variant
descriptions from this graph, including one that yields domain-specific
constructs from the HGVS nomenclature.

\noindent
Our experiments show that our methods in comparison with dbSNP, the
authoritative variant database from the NCBI, result in identical HGVS
descriptions for simple variants and more meaningful descriptions for complex
variants, in particular for repeat expansions and contractions.
\end{abstract}

\todo{position ourselves against structural variants}

\section{Introduction}\label{sec:intro}
Insights into a person's DNA are of relevance for a wide range of applications,
not limited to clinical and research settings. Sequencing is the process to
obtain the genetic sequence from DNA. Typically, this results in segments of
chromosomes that need to be aligned to a reference genome, e.g., Genome
Reference Consortium Human Build 38 (GRCh38) for human samples, in order to
reconstruct the individual's complete chromosomal sequences.

Because the human genome is large~(circa~$3 \cdot 10^9$ nucleotides), and the
individual differences between two genomes are relatively
small~(circa~$0.6\%$~\cite{variation}), it is practical to only look at the
differences.
These differences are called genetic variants. In practice, these are recorded
by most variant calling tools in the Variant Call Format~(VCF)~\cite{vcf}.
Variants appear in various formats in clinical communication, scientific
literature, and are also stored in databases.
The effectiveness of this information exchange largely depends on the specific
format and quality of how these variants are described~\cite{designation,
suggested_nom}.

The Human Genome Variation Society~(HGVS) Variant Nomenclature
Committee~\cite{hgvs} publishes recommendations~\cite{hgvs2024} for HGVS
variant descriptions, that are intended to be human-readable, but provides no
guidelines on how to construct them from sequences.
A standardization effort with a machine-readable focus is the Variant
Representation Specification~(VRS)~\cite{vrs}.
SPDI features a single replacement notation (Sequence, Position, Deletion,
Insertion) and employs the Variant Overprecision Correction Algorithm (VOCA)
for normalization.
The Single Nucleotide Polymorphism Database~(dbSNP)~\cite{dbsnp2024} is a
public archive for genetic variation made available by National Center for
Biotechnology Information (NCBI). The widespread usage of the dbSNP database
makes their specific choice in formatting of HGVS variants quite prevalent.

In~\cite{variant_name}, significant inconsistencies in variant descriptions
across tools and databases are observed, noting that it specifically hinders
automatic lookup.
Often variant description languages define a \emph{normalization} procedure
to select one universally accepted description from a set of possibilities.
Beyond normalization, in~\cite{curation} the authors also discuss many other
challenges to effectively curate variants from literature.
The authors of~\cite{litvar} show that searching for a single description in
PubMed does not retrieve all relevant articles for a variant.
In a more clinical study~\cite{twoincis} the authors observe that differences
in nomenclature can ``hide clinically relevant information in plain sight''.
Challenges with VCF normalization have been addressed in
\cite{improved_vcf,unified}.
In~\cite{deletions} the authors observe that in repetitive genomic regions,
different descriptions of the same deletions can lead to the same result.
A sequence-based normalization method specifically for concise HGVS
descriptions is discussed in~\cite{vde}.
The problem of choosing an accurate variant description and assisting the user
with variant normalization has given rise to many
platforms~\cite{validator2018,mutalyzer,varsome,tmvar3}.

\begin{figure}[ht]
\begin{center}
\resizebox{.9\textwidth}{!}{%
\begin{tikzpicture}
\node[draw=highlight, vertexsm] (a) at (0, 0) {};
\node[vertexsm] (b) at (3, 1.2) {};
\node[draw=highlight, vertexsm] (c) at (3, -1.2) {};
\node[vertexsm] (d) at (6, 0) {};
\node[vertexsm] (e) at (6, 2.4) {};
\node[draw=highlight, vertexsm] (f) at (9, 0) {};
\node[draw=highlight, double, vertexsm] (g) at (12, 0) {};

\path[edge]
    (-1.2, 0) edge[draw=highlight, -{Stealth[fill=highlight, scale=1.2]}] (a)
    (a) edge node[xshift=-.5cm, above] {\texttt{1\_2insTT}} (b)
    (a) edge[draw=highlight, -{Stealth[fill=highlight, scale=1.2]}] node[text=highlight, xshift=-.2cm, below] {\texttt{2C>T}} (c)
    (b) edge node[xshift=.2cm, above] {\texttt{3C>T}} (d)
    (b) edge node[xshift=-.2cm, above] {\texttt{3del}} (e)
    (c) edge node[xshift=-.3cm, above] {\texttt{3\_4insT}} (d)
    (c) edge[draw=highlight, -{Stealth[fill=highlight, scale=1.2]}, bend right=20] node[text=highlight, below] {\texttt{5G>T}} (f)
    (d) edge node[xshift=-.1cm, above] {\texttt{5del}} (f)
    (e) edge node[xshift=.3cm, above] {\texttt{5G>T}} (f)
    (f) edge[bend left=20] node[above] {\texttt{7\_8insT}} (g)
    (f) edge[draw=highlight, -{Stealth[fill=highlight, scale=1.2]}, bend right=20] node[text=highlight, below] {\texttt{8\_9insT}} (g)
;

\end{tikzpicture}%
}
\caption{
A graph containing all LCS alignments for the strings \texttt{ACCTGACT} and
\texttt{ATCTTACTT}. The paths through the graph correspond with six HGVS
descriptions.
The highlighted path with the description \texttt{[2C>T;5G>T;8\_9insT]}
corresponds to the HGVS recommendations.
}
\label{fig:simple-graph}
\end{center}
\end{figure}
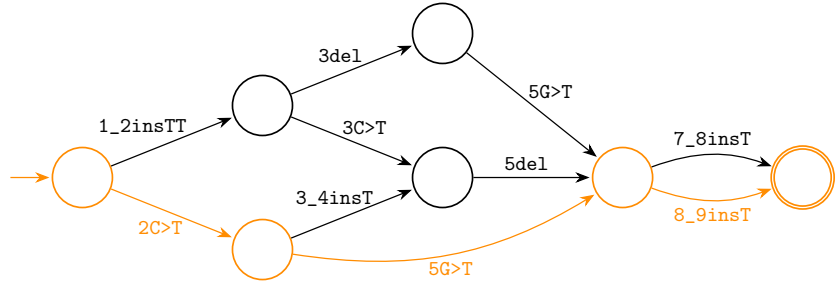

\bigskip
\noindent
In the remainder of this paper we focus on a formal approach to
calculate variant descriptions.
Central is the problem of computing the distance between two strings and obtaining a
series of edit operations~\cite{string_to_string}.
Instead of focusing on \emph{one} series of edit operations, we consider
\emph{all} minimal alignments corresponding to the simple edit
distance~\cite{algebra} to generate variant descriptions.
We introduce a new type of graph that is constructed from all these minimal
alignments; a simplified example visualization using HGVS notation is shown in
Figure~\ref{fig:simple-graph}. Based on the structure of this graph, we provide
three complementary methods to derive variant descriptions. We
experimentally assess the effectiveness of our method by comparing our results
to the HGVS descriptions in dbSNP. Finally, we discuss the broader implications of
our methods and provide forward-looking statements.

\todo{tweak vertical space}

\section{Algorithms and Data Structures}\label{sec:methods}

We start with preliminary definitions that we build upon later.
Let $\Sigma$ be a non-empty finite \emph{alphabet} of \emph{symbols},
e.g., $\Sigma = \{\texttt{A}, \texttt{C}, \texttt{G}, \texttt{T}\}$, and
$S$ a finite \emph{string} over $\Sigma$. The \emph{length} of
string~$S$, written as $|S|$, is the number of symbols in~$S$. The empty
string (without any symbols, also said with length~$0$) is denoted
as~$\varepsilon$. We refer to the symbol at position~$i$ of~$S$ as $S_i$,
with $0 \le i < |S|$. We use the natural extension of this notation to
refer to \emph{substrings} of~$S$, i.e., $S_{i\ldots j}$ is the string
containing the contiguous symbols~$S_i,\ldots,S_{j-1}$,
with~$0 \le i \le j \le |S|$. Note that $S_{i\ldots i} = \varepsilon$.
String~$U$ is a \emph{subsequence} of string~$S$ if $U = \varepsilon$ or
there exists a sequence of integers $i_1,\ldots,i_{|U|}$ such that
$0 \le i_1 < \ldots < i_{|U|} < |S|$ and $U = S_{i_1}\ldots S_{i_{|U|}}$.

To describe differences between two
strings~$R$ (called \emph{reference} sequence) and~$O$ (called
\emph{observed} sequence) we introduce the \emph{replacement} (also known
as substitution) operator~$i\!:\!j\,/\,S$, with~$0 \le i \le j \le |R|$
and~$S$ a substring of~$O$ that replaces the
symbols~$R_i\ldots R_{j - 1}$ with the string~$S$. When $i = j$, the
string~$S$ is inserted immediately before (the symbol on) the $i$-th~position. We denote
the empty replacement as $\lambda$. The string~$R$ can be transformed
into~$O$ by applying a series of replacement operations.

A \emph{longest common subsequence}~(LCS) for two strings~$R$ and~$O$ is
a longest string~$U$ such that $U$ is a subsequence of both~$R$ and~$O$.
In general, many of such LCSs exist for a given $R$ and $O$. Often, only the
length of the LCSs is of interest.
The classic approach to calculate the (length of the) LCSs is to solve the
well-known recurrence relation using dynamic programming to fill a table row by
row~\cite{string_to_string}.

\subsection{All LCS Alignments}\label{sec:all-lcs-alignments}

We are interested in finding all LCS alignments for a pair of highly
similar strings.
It is well-known~\cite{survey} that LCS alignment is related to calculating
the \emph{simple edit distance}, e.g.,~\cite{algebra}.
For only calculating the simple edit distance, the
OND~algorithm~\cite{lcs_myers} and its later refinement the
ONP~algorithm~\cite{lcs_wu} are considered practical methods in both space
and time complexity~\cite{survey}.  However, in general, these methods do not
yield all minimal alignments.
In this section we show that a straightforward extension of the ONP~algorithm
does yield all minimal alignments, providing us with an elegant solution to the
problem.

Contrary to the classic dynamic programming approach that computes a table row by
row, here the calculation iteratively expands along the diagonals of the table.
The first iteration starts with a set of numbered diagonals~$\mathit{FP} = \{0,
\ldots, |R| - |O|\}$, where $0$ refers to the main diagonal.
Each iteration, the diagonals in $\mathit{FP}$ are processed outside in, finishing with
the $|R| - |O|$~diagonal.
Every diagonal is expanded to the maximum of the furthest points of its two flanking diagonals
and possibly further as long as symbols from the two strings match.
As long as the bottom-right of the table is not reached, we add the two flanking
diagonals to the set~$\mathit{FP}$ for the next iteration.
The computed elements are never stored, only keeping track of the contour of
the computed elements by storing the farthest position on the
diagonals is required.

The length of the LCSs for string~$R$ and string~$O$ at position~$(i, j)$, with
$0 \le i < |R|$ and $0 \le j < |O|$, can be defined as:
\begin{equation}\label{eq:lcs}
|\text{LCS}(R_{0\ldots i}, O_{0\ldots j})| = \left\lfloor \frac{i + j - ||R| - |O|| - 2p + |(|R| - i) - (|O| - j)|}{2}\right\rfloor + 1,
\end{equation}
for the $p$-th~iteration from the ONP~algorithm.

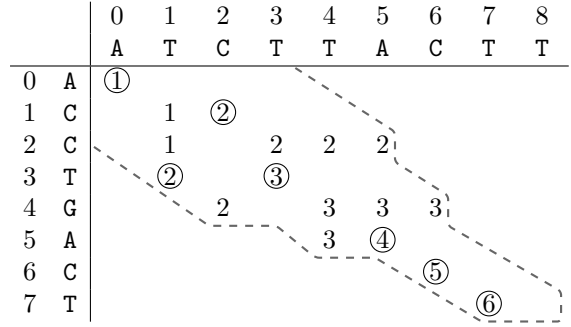
\begin{figure}[ht]
\[
\begin{array}{rc|ccccccccccc}
 & & 0 & 1 & 2 & 3 & 4 & 5 & 6 & 7 & 8 \\
 & & \texttt{A} & \texttt{T} & \texttt{C} & \texttt{T} & \texttt{T} & \texttt{A} & \texttt{C} & \texttt{T} & \texttt{T} \\
\hline
0 & \texttt{A} & \tikzcircled{1} & & & \tikzmark{wu9} & \quad & \quad & \quad & \quad & \quad \\
1 & \texttt{C} & & 1 & \tikzcircled{2} \\
2 & \texttt{C} & & 1 & & 2 & 2 & \tikzmark{wu8}2 \\
3 & \texttt{T} & & \tikzmark{wu0}\tikzcircled{2} & & \tikzcircled{3} \\
4 & \texttt{G} & & & \tikzmark{wu1}2 & \tikzmark{wu2} & 3 & 3 & \tikzmark{wu7}3 \\
5 & \texttt{A} & & & & & \tikzmark{wu3}3 & \tikzmark{wu4}\tikzcircled{4} \\
6 & \texttt{C} & & & & & & & \tikzcircled{5} \\
7 & \texttt{T} & \quad & \quad & \quad & \quad & \quad & \quad & \quad & \tikzmark{wu5}\tikzcircled{6} & \tikzmark{wu6} \\
\end{array}
\]
\begin{tikzpicture}[overlay, remember picture]
\draw[contour] plot coordinates{
    ([xshift=-.1cm] wu0.south) ([xshift=-.1cm] wu1.south) (wu2.south)
    ([xshift=-.1cm] wu3.south) ([xshift=-.1cm] wu4.south)
    ([xshift=-.1cm] wu5.south) ([xshift=.25cm] wu6.south)
    ([xshift=.25cm, yshift=.4cm] wu6.south)
    ([xshift=.25cm] wu7.south) ([xshift=.25cm, yshift=.4cm] wu7.south)
    ([xshift=.25cm] wu8.south) ([xshift=.25cm, yshift=.4cm] wu8.south)
    ([xshift=.25cm, yshift=.4cm] wu9.south)
};
\end{tikzpicture}
\caption{The computed elements by the original ONP~algorithm with their values according to Equation~\ref{eq:lcs}
for $R = \texttt{ACCTGACT}$ and $O = \texttt{ATCTTACTT}$.
The dashed line marks the contour of the solution space, the final set~$\mathit{FP}$.
Matching symbols are circled.
}
\label{fig:wu}
\end{figure}

Figure~\ref{fig:wu} depicts the computed elements by the ONP~algorithm for
$R = \texttt{ACCTGACT}$ and $O = \texttt{ATCTTACTT}$. 
We observe that the contour corresponds to the frontier of the A*
implementation in~\cite{algebra}.
Not all elements in the contour are calculated, including some of the matches.
In order to calculate all LCS alignments, we require all elements with
matching symbols within the contour.
The matching~\texttt{T} at $(3, 4)$ is a concrete example of a match that the
ONP~algorithm does not compute, but that we require.
Therefore, instead of just taking the maximum position of the adjacent diagonals, we
have to check for matching symbols up to the maximum, resulting in Figure~\ref{fig:edit}.

\begin{figure}[ht]
\[
\begin{array}{rc|ccccccccccc}
 & & 0 & 1 & 2 & 3 & 4 & 5 & 6 & 7 & 8 \\
 & & \texttt{A} & \texttt{T} & \texttt{C} & \texttt{T} & \texttt{T} & \texttt{A} & \texttt{C} & \texttt{T} & \texttt{T} \\
\hline
0 & \texttt{A} & \tikzcircled{1} & 1 & 1 & \tikzmark{ed9}1 & \quad & \quad & \quad & \quad & \quad \\
1 & \texttt{C} & 1 & 1 & \tikzcircled{2} & 2 & 2 \\
2 & \texttt{C} & 1 & 1 & \tikzcircled[a]{2} & 2 & 2 & \tikzmark{ed8}2 \\
3 & \texttt{T} & & \tikzmark{ed0}\tikzcircled{2} & 2 & \tikzcircled[b]{3} & \tikzcircled{3} & 3 \\
4 & \texttt{G} & & & \tikzmark{ed1}2 & \tikzmark{ed2}3 & 3 & 3 & \tikzmark{ed7}3 \\
5 & \texttt{A} & & & & & \tikzmark{ed3}3 & \tikzmark{ed4}\tikzcircled[c]{4} & 4 \\
6 & \texttt{C} & & & & & & & \tikzcircled{5} & 5 \\
7 & \texttt{T} & \quad & \quad & \quad & \quad & \quad & \quad & \quad & \tikzmark{ed5}\tikzcircled[d]{6} & \tikzmark{ed6}\tikzcircled{6} \\
\end{array}
\]
\begin{tikzpicture}[overlay, remember picture]
\draw[contour] plot coordinates{
    ([xshift=-.1cm] ed0.south) ([xshift=-.1cm] ed1.south) (ed2.south)
    ([xshift=-.1cm] ed3.south) ([xshift=-.1cm] ed4.south)
    ([xshift=-.1cm] ed5.south) ([xshift=.25cm] ed6.south)
    ([xshift=.25cm, yshift=.4cm] ed6.south)
    ([xshift=.25cm] ed7.south) ([xshift=.25cm, yshift=.4cm] ed7.south)
    ([xshift=.25cm] ed8.south) ([xshift=.25cm, yshift=.4cm] ed8.south)
    ([xshift=.25cm, yshift=.4cm] ed9.south)
};
\node[circle, inner sep=0, minimum size=9pt] (a0) at (a) {};
\node[circle, inner sep=0, minimum size=9pt] (b0) at (b) {};
\node[circle, inner sep=0, minimum size=9pt] (c0) at (c) {};
\node[circle, inner sep=0, minimum size=9pt] (d0) at (d) {};
\draw
    (tangent cs:node=a0, point={(b0)}) -- (tangent cs:node=b0, point={(a0)}, solution=2)
    (tangent cs:node=a0, point={(b0)}, solution=2) -- (tangent cs:node=b0, point={(a0)})
    (tangent cs:node=c0, point={(d0)}) -- (tangent cs:node=d0, point={(c0)}, solution=2)
    (tangent cs:node=c0, point={(d0)}, solution=2) -- (tangent cs:node=d0, point={(c0)})
;
\end{tikzpicture}
\caption{The computed elements of our extension of the ONP~algorithm with their values according to Equation~\ref{eq:lcs}
for $R = \texttt{ACCTGACT}$ and $O = \texttt{ATCTTACTT}$.
Consecutive matches are grouped.
}
\label{fig:edit}
\end{figure}

\todo{lcs-nodes name consistently}

\noindent
As we are interested in all LCS alignments we also require an
efficient method for storing all the matches.
Given the context of highly similar sequences, we record all matches into
groups of consecutive matches, represented by a triple $(i, j, \ell)$, where
$i$ and $j$ are the coordinates of
the positions of the first match together with $\ell$, the number of consecutive
matches, effectively limiting the number of entries.
The triples are stored in an \emph{LCS-nodes} data structure indexed on the
zero-based string position in an LCS of the last match of the node.
Matches $(0, 0)$, $(1, 2)$, $(2, 2)$, $(3, 1)$, $(3, 3)$, $(3, 4)$, $(5, 5)$,
$(6, 6)$, $(7, 7)$, $(7, 8)$ from Figure~\ref{fig:edit} are thus stored as:

\begin{align*}
0&: (0, 0, 1) \\
1&: (1, 2, 1), (3, 1, 1) \\
2&: (2, 2, 2), (3, 4, 1) \\
3&: \lambda \\
4&: \lambda \\
5&: (5, 5, 3), (7, 8, 1), \\
\end{align*}
\noindent
where $\lambda$ denotes the empty list. The number of consecutive matches
$\ell$ is also referred to as the length of an element.

The ONP~algorithm swaps $R$ and $O$ when $|R| < |O|$ as it explores the problem from the
perspective of deletions. It can freely do so as it only calculates the distance.
As we compute the edit operations from $R$ to $O$, our extension takes care of not
having to swap the strings.

\subsection{Constructing the Compressed LCS-graph}\label{sec:clcs-graph}

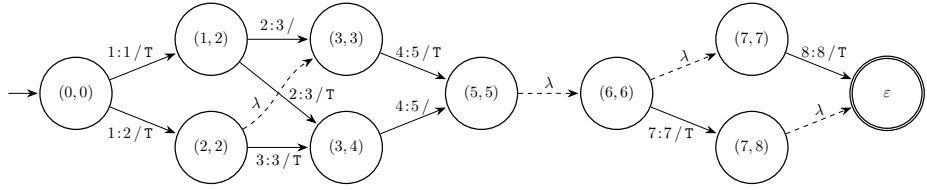
\begin{figure}[ht]
\resizebox{\textwidth}{!}{%
\begin{tikzpicture}
\node[vertex] (a) at (0, 0) {$(0, 0)$};
\node[vertex] (b) at (3, 1.2) {$(1, 2)$};
\node[vertex] (c) at (3, -1.2) {$(2, 2)$};
\node[vertex] (d) at (6, 1.2) {$(3, 3)$};
\node[vertex] (e) at (6, -1.2) {$(3, 4)$};
\node[vertex] (f) at (9, 0) {$(5, 5)$};
\node[vertex] (g) at (12, 0) {$(6, 6)$};
\node[vertex] (h) at (15, 1.2) {$(7, 7)$};
\node[vertex] (i) at (15, -1.2) {$(7, 8)$};
\node[double, vertex] (j) at (18, 0) {$\varepsilon$};

\path[lambda edge]
    (c) edge[out=30, in=210] node[xshift=-.5cm, yshift=-.2cm] {$\lambda$} (d)
    (f) edge node[above] {$\lambda$} (g)
    (g) edge node[above] {$\lambda$} (h)
    (i) edge node[above] {$\lambda$} (j)
;

\path[edge]
    (-1.5, 0) edge (a)
    (a) edge node[xshift=-.3cm, above] {\variant{1}{1}{T}} (b)
    (a) edge node[xshift=-.3cm, below] {\variant{1}{2}{T}} (c)
    (b) edge node[above] {\variant{2}{3}{}} (d)
    (c) edge node[below] {\variant{3}{3}{T}} (e)
    (b) edge node[near end, xshift=.3cm, above] {\variant{2}{3}{T}} (e)
    (d) edge node[xshift=.1cm, above] {\variant{4}{5}{T}} (f)
    (e) edge node[above] {\variant{4}{5}{}} (f)
    (h) edge node[xshift=.1cm, above] {\variant{8}{8}{T}} (j)
    (g) edge node[xshift=-.3cm, below] {\variant{7}{7}{T}} (i)
;
\end{tikzpicture}%
}
\caption{
LCS-graph of $R = \texttt{ACCTGACT}$ and $O = \texttt{ATCTTACTT}$.
Nodes represent matches and are labeled with their respective positions~$(i, j)$.
Edges describe the minimal replacements between the nodes.
Dashed edges represent empty replacements.
The leading edge indicates the source node and the double-circled node
indicates the sink node.
}
\label{fig:lcs-graph}
\end{figure}

The LCS-graph presented in~\cite{efficient_alllcs,algebra} is a directed acyclic graph
that consists of nodes representing single symbol matches for all LCSs.
These graphs always have a single source node and a single sink node.
Labeled edges connect nodes for consecutive symbols in an LCS.
Figure~\ref{fig:lcs-graph} shows the LCS-graph for the earlier discussed
example.

\begin{figure}[hb]
\resizebox{\textwidth}{!}{%
\begin{tikzpicture}
\node[vertex] (s6) at (0, 0) {$(0, 0, 1)$};
\node[vertex] (s4) at (3, 1.2) {$(1, 2, 1)$};
\node[vertex] (s5) at (3, -1.2) {$(2, 2, 1)$};
\node[vertex] (s2) at (6, 1.2) {$(3, 3, 1)$};
\node[vertex] (s3) at (6, -1.2) {$(3, 4, 1)$};
\node[vertex] (s1) at (9, 0) {$(5, 5, 3)$};
\node[double, vertex] (s0) at (12, 0) {$(7, 8, 1)$};

\path[lambda edge]
    (s5) edge[out=30, in=210] node[xshift=-.5cm, yshift=-.2cm] {$\lambda$} (s2)
;

\path[edge]
    (-1.5, 0) edge (s6)
    (s6) edge node[xshift=-.3cm, above] {\variant{1}{1}{T}} (s4)
    (s6) edge node[xshift=-.3cm, below] {\variant{1}{2}{T}} (s5)
    (s4) edge node[above] {\variant{2}{3}{}} (s2)
    (s5) edge node[below] {\variant{3}{3}{T}} (s3)
    (s4) edge node[near end, xshift=.3cm, above] {\variant{2}{3}{T}} (s3)
    (s2) edge node[xshift=.1cm, above] {\variant{4}{5}{T}} (s1)
    (s3) edge node[above] {\variant{4}{5}{}} (s1)
    (s1) edge[bend left=20] node[above] {\variant{7}{7}{T}} (s0)
    (s1) edge[bend right=20] node[below] {\variant{8}{8}{T}} (s0)
;
\end{tikzpicture}%
}

\caption{
Compressed LCS-graph of $R = \texttt{ACCTGACT}$ and $O = \texttt{ATCTTACTT}$.
Nodes represent matches and are labeled as $(i, j, \ell)$.
Edges describe the minimal replacements between the nodes.
Multiple edges representing unique minimal replacements between two nodes can exist.
}
\label{fig:clcs-graph}
\end{figure}
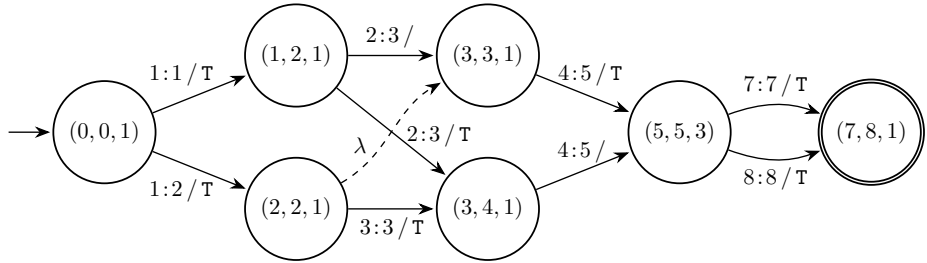

In this section we introduce the \emph{compressed LCS-graph}~(cLCS-graph) that is
a directed acyclic multigraph constructed from the (consecutive) matches in the
LCS-nodes data structure. Figure~\ref{fig:clcs-graph} shows the cLCS-graph for
the earlier discussed example.

\todo{move this graph further down?}

We say that node~$(i, j, k)$ is \emph{dominated} by node~$(i', j', k')$ if
$i + k < i' + k'$ and $j + k < j' + k'$.
There is an edge between node~$(i, j, \ell)$ and node~$(i', j', \ell')$
for every $0 \le k < \ell$ and every $0 \le k' < \ell'$ such that the following
two conditions are met:
node~$(i, j, k)$ is dominated by node~$(i', j', k')$ and
they are consecutive matches in an LCS:
$|\text{LCS}(R_{0\ldots i + k}, O_{0\ldots j + k})| = |\text{LCS}(R_{0\ldots i'
+k'}, O_{0\ldots j' + k'})| - 1$.

The edges are labeled with the replacement: $i + k + 1 : i' + k' /
O_{j + k + 1 \ldots j' + k' - 1}$.
For example, in Figure~\ref{fig:clcs-graph} there is an edge between nodes
$(5, 5, 3)$ and $(7, 8, 1)$ labeled \variant{7}{7}{T} with $k = 1$ and $k' = 0$.
Note that we do not only compress nodes; we also compress edges when all
minimal replacements between two nodes are present, e.g., \variant{1}{2}{T} represents
the two different alignments [\variant{1}{1}{T}, \variant{1}{2}{}] and
[\variant{1}{2}{}, \variant{2}{2}{T}].

\todo{quantify lcs compression between two graphs}
\todo{quantify the alignments that clcs represents}

We can systematically and efficiently construct the graph using the LCS-nodes
data structure. Processing the elements in the LCS-nodes is based on their
index from high to low. We check the elements at an indexed LCS position from
right to left, as they are conveniently ordered as a direct consequence of
when they were discovered during the alignment.
For an element with length~$\ell$ at index~$p$, we check the conditions
with all other elements at indices in~${p, p-1, \ldots, p - \ell}$.
When the conditions are met, the corresponding edges are added to the graph.
The current element is then discarded from the LCS-nodes data structure.
In situations when there is no match at the beginning and/or at the end of the
sequences, a respective source and/or sink node with an $\epsilon$ match is
added to the graph, so that also these graphs have a unique source and sink
node.

Not all processed matches are part of the ultimate alignments. Because of the
processing order these elements can be easily recognized, as they have no
path to the sink node, and are therefore not added to the graph.
For example, element $(3, 1, 1)$ from Figure~\ref{fig:edit} is not part of any LCS
alignment. In general, these elements are also computed by the ONP~algorithm (see
Figure~\ref{fig:wu}).

\subsubsection{Splitting Matches}

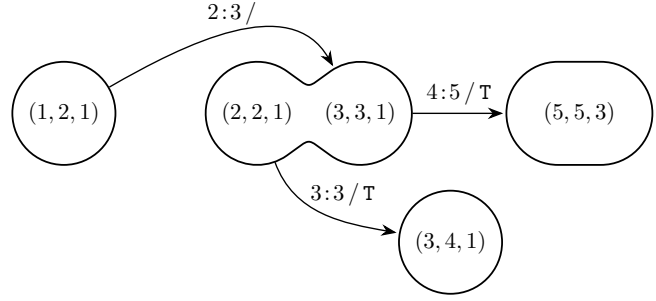
\begin{figure}[ht]
\begin{center}
\resizebox{0.7\textwidth}{!}{%
\begin{tikzpicture}
\node[vertex] (a) at (-3, 0) {$(1, 2, 1)$};
\node[circle, minimum size=1.6cm] (b1) at (0, 0) {$(2, 2, 1)$};
\node[circle, minimum size=1.6cm] (b2) at (1.6, 0) {$(3, 3, 1)$};
\node[draw, minimum height=1.6cm, minimum width=2.5cm, rounded rectangle, thick] (c) at (5, 0) {$(5, 5, 3)$};
\node[vertex] (d) at (3, -2) {$(3, 4, 1)$};

\draw[thick]
    ([shift=(55:.8)] 0, 0) arc (55:305:.8)
    .. controls ++(35:.5) and ++(145:.5) ..
    ([shift=(235:.8)] 1.6, 0) arc (235:485:.8)
    .. controls ++(35:-.5) and ++(145:-.5) ..
    ([shift=(55:.8)] 0, 0)
;

\path[edge]
    (a) edge[out=30, in=120] node[above] {\variant{2}{3}{}} (b2)
    (b1) edge[out=290, in=170] node[xshift=.3cm, above] {\variant{3}{3}{T}} (d)
    (b2) edge node[above] {\variant{4}{5}{T}} (c.west)
;
\end{tikzpicture}%
}
\caption{The immediate neighborhood of node~$(2, 2, 2)$ from
Figure~\ref{fig:clcs-graph}. Node~$(2, 2, 2)$ is split into two nodes~$(2, 2,
1)$ and $(3, 3, 1)$ in order to avoid path~$[$\variant{2}{3}{},
\variant{3}{3}{T}$]$.}\label{fig:split}
\end{center}
\end{figure}

As a consequence of compressing the consecutive matches, we run the risk of
introducing ``inverse'' paths (alignments) that are not present in the
classic LCS-graph. These inversions are caused by incoming edges that enter
compressed nodes ``beyond'' outgoing edges. In Figure~\ref{fig:split} we
illustrate the situation occurring with element $(2, 2, 2)$ from
the cLCS-graph in Figure~\ref{fig:edit}.

\noindent
Element~$(2, 2, 2)$ represents two consecutive matches, present as separate nodes $(2, 2)$
and~$(3, 3)$ in the LCS-graph. Without countermeasures, the
path~$[$\variant{2}{3}{}, \variant{3}{3}{T}$]$ would be present. However, this path does not exist
in the LCS-graph because of the explicit (unlabeled) edge between $(2, 2)$ and $(3, 3)$.
To remedy this situation, we create a new node by splitting the conflict-inducing
node immediately before the conflicting incoming edge and connect these two nodes
with an unlabeled ($\lambda$) edge. For the remainder of the calculation only the
first (newly created) node needs to be considered.

\section{Variant Description Extraction}

\todo{define variant}

The cLCS-graph introduced in~Section~\ref{sec:methods} contains all relevant
changes for a genetic variant and could therefore be considered a complete
representation of that variant, often corresponding to multiple variant
descriptions.
In contrast, in the domain one often needs to choose a single textual variant
description for practical purposes.
In order to obtain a variant description from a cLCS-graph, we define the
problem of extraction: to calculate a route from source to sink allowing to use
both existing edges and additional shortcut replacements between two nodes
that are not directly connected, where the shortcuts do not introduce
cycles in the graph. Shortcuts enable summarization of (potentially) complex
parts of the graph, but do not represent an LCS alignment.

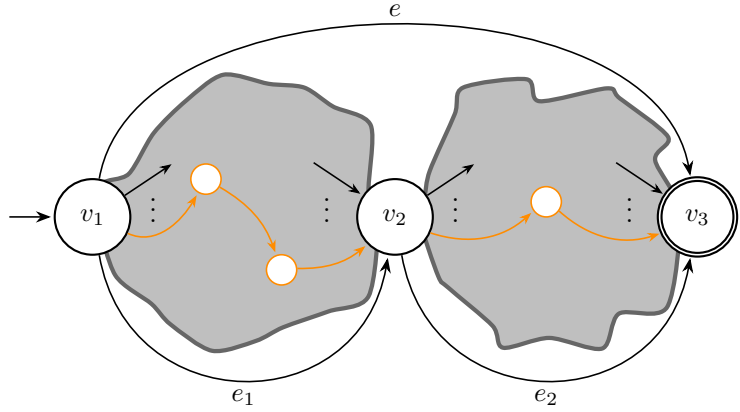
\begin{figure}[ht]
\begin{center}
\begin{tikzpicture}
\draw[draw=black!60, fill=black!25, ultra thick]
plot[smooth cycle] coordinates{
    (1.78, -0.02)
    (1.65, 0.32)
    (1.68, 1.21)
    (1.18, 1.46)
    (0.80, 1.71)
    (0.33, 1.76)
    (-0.27, 1.88)
    (-0.76, 1.54)
    (-1.19, 1.11)
    (-1.52, 0.57)
    (-1.88, 0.43)
    (-1.92, -0.02)
    (-1.70, -0.77)
    (-1.22, -1.15)
    (-0.97, -1.42)
    (-0.42, -1.77)
    (0.16, -1.64)
    (0.52, -1.46)
    (1.12, -1.25)
    (1.65, -1.05)
}
plot[smooth cycle] coordinates{
    (5.76, 0.06)
    (5.44, 0.75)
    (5.61, 1.06)
    (4.97, 1.57)
    (4.80, 1.70)
    (3.91, 1.53)
    (3.74, 1.82)
    (2.90, 1.71)
    (2.79, 1.36)
    (2.46, 0.61)
    (2.39, 0.25)
    (2.33, -0.04)
    (2.57, -0.78)
    (2.71, -0.94)
    (3.29, -1.44)
    (3.43, -1.73)
    (4.04, -1.60)
    (4.74, -1.74)
    (4.96, -1.28)
    (5.60, -1.19)
}
;

\node[circle, draw, fill=white, minimum size=1cm, thick] (a) at (-2, 0) {$v_1$};
\node[circle, draw, fill=white, minimum size=1cm, thick] (b) at (2, 0) {$v_2$};
\node[circle, draw, double, fill=white, minimum size=1cm, thick] (c) at (6, 0) {$v_3$};

\node (a1) at (-.8, .8) {};
\node at (-1.2, .2) {$\vdots$};
\node (a2) at (.8, .8) {};
\node at (1.1, .2) {$\vdots$};

\node (b1) at (3.2, .8) {};
\node at (2.8, .2) {$\vdots$};
\node (b2) at (4.8, .8) {};
\node at (5.1, .2) {$\vdots$};

\node[circle, draw=highlight, fill=white, minimum size=.4cm, semithick] (s) at (-.5, .5) {};
\node[circle, draw=highlight, fill=white, minimum size=.4cm, semithick] (t) at (.5, -.7) {};
\node[circle, draw=highlight, fill=white, minimum size=.4cm, semithick] (u) at (4, .2) {};

\path[edge, -{Stealth[scale=1]}]
    (-3.1, 0) edge (a)
    (a) edge[out=80, in=100, looseness=.9] node[above] {$e$} (c)
    (a) edge[out=-80, in=-100, looseness=1.5] node[below] {$e_1$} (b)
    (b) edge[out=-80, in=-100, looseness=1.5] node[below] {$e_2$} (c)
;

\path[edge, -{Stealth[scale=.8]}]
    (a) edge (a1)
    (a2) edge (b)
    (b) edge (b1)
    (b2) edge (c)
;

\path[draw=highlight, edge, fill=highlight, -{Stealth[scale=.8]}]
    (a) edge[out=-25, in=-120] (s)
    (s) edge[bend left=20] (t)
    (t) edge[bend right=20] (b)
    (b) edge[bend right=30] (u)
    (u) edge[bend right=30] (c)
;
\end{tikzpicture}
\end{center}
\caption{A schematic view of an instance of a cLCS-graph with the supremal variant, all-encompassing
single replacement $e$. All paths go through nodes $v_1$, $v_2$, and $v_3$,
partitioning the graph in local supremal parts described by replacements $e_1$ and $e_2$.
The highlighted path is the canonical path.}
\label{fig:hierarchy}
\end{figure}

Within the domain, the ultimate goal regarding variant descriptions, is to
determine a single variant description out of the numerous ways of describing
that variant, often beyond considering only LCS alignments.
In this section we present three alternative extraction methods specifically
tailored toward yielding descriptions with relevance in the domain.
Figure~\ref{fig:hierarchy} gives a schematic overview of the three methods and
their relationships:
i) the \emph{supremal variant}, which is especially suited for describing
small deletion-insertions analogous to SPDI with VOCA normalization.
It is a single unique all-encompassing replacement ($e$ in Figure~\ref{fig:hierarchy});
ii) the \emph{local supremal variant} that naturally extends the supremal
variant to a version that caters for allele descriptions (multiple small
deletion-insertions \emph{in cis}) ($e_1$ and $e_2$ in Figure~\ref{fig:hierarchy});
and iii) the \emph{canonical variant}, which is specifically geared toward variant
reporting in the domain, e.g., using HGVS (the highlighted path in
Figure~\ref{fig:hierarchy}). The NCBI uses the term ``Canonical SPDI'' for
similar purposes, but note that these are not necessarily the same.

\subsection{Supremal Variants}\label{sec:supremal}

Every variant can be described by the complete deletion of $R$ and the
subsequent insertion of $O$.
Stretching that argument to the extreme, the variant can thus also be described
by only $O$ (with an empty $R$). For small variants on large sequences this is
generally not practical.

The supremal variant is a similar, but generally more compact, description of
a single replacement that captures all relevant changes. It is also used for
calculating the relation between two variants as introduced in~\cite{algebra}.
In this paper we restrict ourselves to the changes within the supremal
variant for determining variant descriptions.
Using the cLCS-graph, the supremal variant is constructed as a single
replacement between the source and sink nodes, by taking the
position of the first outgoing edge of the source node and the position of the
last incoming edge of the sink node. For the example in Figure~\ref{fig:clcs-graph},
the supremal variant is \variant{1}{8}{TCTTACTT}.

In practice, variants are small changes with regard to a large reference
sequence, leading to many consecutive matches in both the source and sink
nodes. Most of these matches are outside of the supremal variant. For
convenience we trim the source and sink nodes to only include the matches that are
inside the supremal variant. For the example in Figure~\ref{fig:clcs-graph}
this leads to relabeling of the source node from $(0, 0, 1)$ to $(1, 1, 0)$.
The sink remains unchanged as the matching \texttt{T} at $(7, 8)$ is inside the
supremal variant.
Note that after trimming, the supremal variant can be derived directly from the
source and the sink node labels without looking at the edges.

Conversely, a supremal variant description is the minimally required input for
constructing the cLCS-graph. In practice this makes the supremal variant
description a good candidate for storing variants.
VOCA used in SPDI and VRS attempts to achieve a similar goal by
enlarging the variant. However, this often yields a smaller deletion-insertion
than the supremal variant.

Even without initial access to the supremal variant, it often pays off to
repeatedly estimate the supremal variant when constructing the cLCS-graph and
checking whether the estimation is correct, by determining if matches are
trimmed at both the source and the sink. If not, the estimate should be
enlarged.

\subsection{Local Supremal Variants}\label{sec:local}

In the domain it becomes more common to look at combinations of multiple
changes. These combinations on the same molecule are said to be phased,
\emph{in cis} or part of the same allele and can be written down as phase sets,
e.g., in VCF, or allele descriptions, e.g., in HGVS.

The technique from Section~\ref{sec:supremal} properly calculates the supremal
variant for alleles too. Despite being technically correct, in many cases it
is not a convenient way to describe such combinations of changes as they can
result in very large deletion-insertions. Here we use a specific property of the
cLCS-graph to come to more meaningful descriptions.

In Figure~\ref{fig:clcs-graph}, we identify three nodes in the cLCS-graph that
are part of every path from the source node to the sink node:
$(0, 0, 1)$, $(5, 5, 3)$, and $(7, 8, 1)$.
All paths have at least one common match in each of these nodes, with the
possible exception of zero-length source and sink nodes.
We argue to partition the graph using these common matches.

The nodes we have identified are called \emph{dominators}~\cite{dominance} of the sink node in
the field of control-flow analysis~\cite{flow-diagrams}. Note that these
dominators should not be confused with domination used in the construction of
the cLCS-graph in Section~\ref{sec:clcs-graph}.
Analogous to the definition of dominators, we define the set of
\emph{post-dominators} of a node $n$ as:
\begin{equation}\label{eq:postdom}
\mathrm{PostDom}(n) = \begin{cases} \left \{ n \right \} & \mathrm{if}\,n = n_0, \\
\left \{ n \right \} \cup \bigcap_{s \in \mathrm{successors}(n)}
\mathrm{PostDom}(s) & \mathrm{otherwise}. \end{cases}
\end{equation}
As our graph has a single sink node, finding the post-dominators of the
source node results in the same nodes as finding the dominators of the sink
node.
In implementations where only outgoing edges are stored, post-dominators can be
calculated more efficiently.

An efficient algorithm applies Equation~\ref{eq:postdom} for each node~$n$ in a single
post-order traversal, while additionally keeping track of the maximal position
for each incoming edge per node.
The resulting set of post-dominators for the source node
$\{(0, 0, 1), (5, 5, 3), (7, 8, 1)\}$ indeed consists of the nodes that are part of
every path in the graph.
Making use of the maximal position of the incoming edges for each of these
nodes and the minimal positions of their outgoing edges, we construct the
replacements between these nodes analogous to the construction of the supremal
variant.
This results in the local supremal variant with replacements
\variant{1}{5}{TCTT} and \variant{7}{8}{TT}. In HGVS this is described as
\texttt{[2\_5delinsTCTT;8delinsTT]}.

\begin{figure}[ht]

\resizebox{\textwidth}{!}{%
\begin{tikzpicture}
\node[vertex] (s0) at (0, 0) {$(0, 0, 1)$};
\node[vertex] (s1) at (3, 0) {$(2, 1, 1)$};
\node[vertex] (s2) at (5.5, 1.6) {$(1, 3, 3)$};
\node[vertex] (s3) at (8, 0) {$(2, 2, 4)$};
\node[double, vertex] (s4) at (11, 0) {$(5, 8, 1)$};

\path[edge]
    (-1.5, 0) edge (s0)
    (s0) edge node[above] {\variant{1}{2}{}} (s1)
    (s0) edge[out=35, in=180] node[above] {\variant{1}{1}{TT}} (s2)
    (s0) edge[out=-35, in=200] node[below] {\variant{1}{2}{T}} (s3)
    (s1) edge node[above] {\variant{3}{3}{T}} (s3)
    (s3) edge[bend left=15] node[above] {\variant{5}{5}{AGC}} (s4)
    (s3) edge[bend right=15] node[below] {\variant{6}{6}{GCA}} (s4)
    (s2) edge[out=0, in=130] node[above] {\variant{4}{5}{GC}} (s4)
;
\end{tikzpicture}%
}
\caption{The cLCS-graph of $[$\variant{1}{1}{TT}, \variant{4}{5}{GC}$]$ with $R = \texttt{CATATATCG}$.
Other than the source and the sink nodes, there is no node through which all
paths traverse.
}
\label{fig:lcs-composition}
\end{figure}
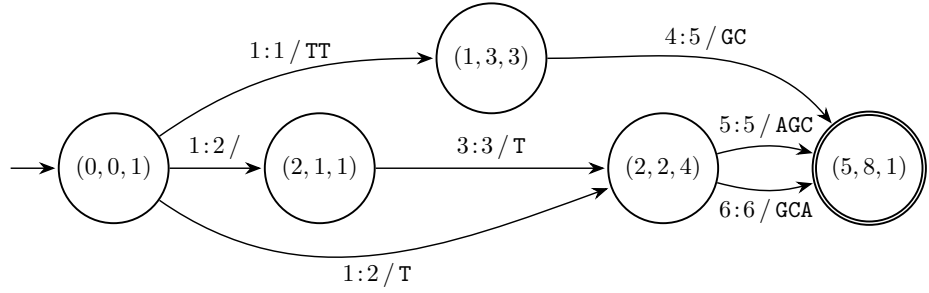

Perhaps counterintuitively, an allele description composed from more variants
that are supremal variants to a reference, does not necessarily lead
to a local supremal variant with those variants as parts. For example,
combining supremal variants \variant{1}{1}{TT} and \variant{4}{5}{GC}, with $R
= \texttt{CATATATCG}$, does not yield the local supremal variant
$[$\variant{1}{1}{TT}, \variant{4}{5}{GC}$]$, but a single replacement
\variant{1}{6}{TTATAGCA} as (local) supremal variant shown in
Figure~\ref{fig:lcs-composition}, emphasizing the need for an extraction method
specifically for variants that consist of multiple parts.

\subsection{Choosing a Canonical Variant}

So far we have used graph properties to find (local) supremal variants, which
generally lead to non-minimal coarse-grained replacements.
However, in the domain the variant descriptions are often more fine-grained.
We introduce a third extraction method based on the well-known shortest path
problem. Intuitively, when presented with a choice of paths through the graph,
shorter paths lead to more concise descriptions.
We consider the subgraph of the cLCS-graph consisting of only the shortest paths.
All edges have weight~$1$ except for unlabeled ($\lambda$) edges which have
weight~$0$.
For the example in Figure~\ref{fig:clcs-graph}, when we consider the part of the
graph between nodes $(0, 0, 1)$ and $(5, 5, 3)$, there is a unique shortest
path of weight~$2$ with edges \variant{1}{2}{T} and \variant{4}{5}{T}.

If there is no unique shortest path, we employ the extraction method
from Section~\ref{sec:local} on the shortest paths subgraph.
In Figure~\ref{fig:clcs-graph} there is no unique shortest path in the part of
the graph between nodes $(5, 5, 3)$ and $(7, 8, 1)$.
The two edges, with weight~$1$, \variant{7}{7}{T} and \variant{8}{8}{T} result
in the replacement \variant{7}{8}{TT}.
Here, these parallel edges indicate the presence of a repeat unit expansion.

\subsection{Constructing HGVS Descriptions}\label{sec:hgvs}

The extraction of the canonical variant results in a series of replacements.
In HGVS more domain-specific constructs exist. Here we translate our
replacements to HGVS descriptions. If for a replacement the deleted or
inserted part is empty, the HGVS counterpart is an \texttt{ins} or
\texttt{del}, respectively, e.g., in Figure~\ref{fig:clcs-graph}, \variant{1}{1}{T} translates to \texttt{1\_2insT} and
\variant{4}{5}{} translates to \texttt{5del}. A replacement where both the
deleted and inserted parts have length $1$ is described in HGVS as a
substitution (also called a single nucleotide variant (SNV) in the domain),
e.g., in Figure~\ref{fig:clcs-graph}, \variant{2}{3}{T} translates to \texttt{3C>T}.

For the remaining replacements, we check if the replacement is a proper repeat
expansion or contraction using a single repeat unit by using the Longest Prefix
Suffix preprocessing step from the Knuth-Morris-Pratt~algorithm~\cite{kmp}.
The identified repeat cases are then described using the HGVS bracketed
repeat syntax, with an exception for a repeat expansion from $1$ to $2$ units,
which is described as a duplication, e.g. in Figure~\ref{fig:clcs-graph}, \variant{7}{8}{TT} translates to
\texttt{8dup}.
Alternatively, if the inserted part of the replacement is the reverse
complement of the deleted part, we describe this as an HGVS inversion.
Otherwise the replacement is described using the HGVS \texttt{delins}
construct.
The HGVS description for the example from Figure~\ref{fig:clcs-graph} is
\texttt{[2C>T;5G>T;8dup]}.

\section{Experiments}\label{sec:experiments}
\todo{mention rsIDs of variants and variants groups}
\todo{properly break long variants}

In order to assess the impact of our method on prevailing variant descriptions,
we compare our extracted HGVS descriptions to descriptions in
dbSNP~\cite{dbsnp2024} Build~156. The dbSNP database is a popular public
resource containing human
single nucleotide variations, microsatellites, and small-scale insertions and
deletions, generally considered the definitive source for variant descriptions.
For all variants on genomic reference sequence NC\_000001.11 (chromosome~1 of
GRCh38), we retrieve both the SPDI and HGVS descriptions.
The SPDI descriptions are used as input to construct the sequences to apply our
method to. Our generated HGVS descriptions are then compared to the respective HGVS
descriptions from dbSNP.
For an indication of performance, our Python implementation is
able to compare about $40$~variants per second per core on modern hardware.

\begin{table}
\centering
\caption{The categorized counts of the textual comparisons between the HGVS entries
in dbSNP and our generated HGVS descriptions.}
\label{table:numbers}
\begin{tabular}{lr}
\toprule
 \multicolumn{2}{@{}l}{\textbf{Identical}}\\
    Substitutions (SNVs) & $91,\!487,\!743$ \\
    Indels & $5,\!374,\!091$ \\
    Repeats & $1,\!278,\!940$ \\
    Inversions & $3,\!374$ \\
  \midrule
     \multicolumn{2}{@{}l}{\textbf{Different}}\\
    Mononucleotide repeats & $3,\!434,\!553$ \\
    Mixed repeats & $317,\!578$ \\
    Inserted sequence repeats & $154,\!866$ \\
    Allele descriptions & $10,\!310$ \\
    Inversions & $26$ \\
  \midrule
   \multicolumn{2}{@{}l}{\textbf{Other}}\\
    Inserted sequence repeats & $18,\!555$ \\
    Ambiguous nucleotides & $668$ \\
  \midrule
    Total & $102,\!080,\!704$ \\
  \bottomrule
\end{tabular}
\end{table}

Table~\ref{table:numbers} shows the results of the comparison of our
descriptions using the method from Section~\ref{sec:hgvs} with the descriptions in dbSNP.
We skipped $18,\!555$~entries because of uninterpretable HGVS descriptions in
dbSNP. We have reported those to the maintainers of the database and they will
hopefully get corrected in a future release.
Additionally, because of the exact nature of our method, we skipped $668$~entries
with ambiguous nucleotide symbols such as~\texttt{N}.

There are $98,\!144,\!148$~identical descriptions, which include all
$91,\!487,\!743$~SNVs present in the database (e.g., \texttt{175292543T>C}).
The remaining identical descriptions are $6,\!656,\!405$ out of the
$10,\!573,\!738$~non-SNVs. These additional identical descriptions include
deletions, e.g., \texttt{206841458del}; insertions,
e.g., \texttt{205259305\_205259306insAG}; simple multinucleotide repeats,
e.g., \texttt{93576231\_\nolinebreak[0]93576234AC[3]}; and inversions, e.g.,
\texttt{195156503\_195156504inv}. Although the simple multinucleotide repeats
slightly differ in syntax from the entries in the database, most likely because of changed
HGVS recommendations regarding repeats, they are semantically identical.

Shifting our attention to the differences, for a total of
$3,\!917,\!333$~variants,
we provide a different description than the entries in dbSNP.
By far the largest fraction of differences consists of mononucleotide repeats.
For a representative example, dbSNP describes a duplication
\texttt{175145557\_175145575dup}, while we describe a repeat expansion
\texttt{175145553\_175145575A[42]}. Similarly, dbSNP describes mononucleotide
repeat contractions as a deletion. When a repeat expansion cannot be expressed
as a duplication (too few repeat units present in the reference), it is
described in dbSNP as an insertion instead. This is surprising, as multinucleotide
repeats are actually described as repeat expansions or contractions in dbSNP.
Furthermore, the normalized SPDI representation does cover all the repeat units
in the reference. Also, in \cite{spdi} an explicit example is given of a
mononucleotide repeat expansion described as such in HGVS. We prefer the
repeat notation for all these cases.

Central to the main message of this paper, $10,\!310$~variants that are described
as a deletion-insertion in dbSNP are described as an allele by us.
For example \texttt{148693179\_148693189delinsGGAAATAAAAC} is
described by us as \texttt{[148693179A>G;148693189G>C]}. Interestingly, both variants
\texttt{148693179A>G} and \texttt{148693189G>C} also appear individually in the
database.

While most of the inversions turn up identical, a small number of inversions in
dbSNP are described by our method as an allele,
e.g., \texttt{221541589\_221541591inv} vs. \texttt{[221541588\_221541589insC;221541591del]}.

Besides describing contractions or expansions of repeat units in the reference
sequence, HGVS also prescribes a repeat-like notation to compress the
inserted sequence. In this experiment there are $154,\!866$~inserted repeated
sequences in dbSNP, also including mononucleotide repeats, that are described
by our method as a repeated insertion, e.g.,
\texttt{149733078delinsTGTTTTGTTTTGTTTTGTTT} vs.
\texttt{149733078delinsTGTTT[4]}.

The last category to address consists of mixed repeats. In the dataset from dbSNP
there are $317,\!578$~entries that describe variants composed of multiple
repeat units and their respective repeat count,
e.g., \texttt{155164866\_155164881T[23]\nolinebreak[0]GTTTTTTTTT[2]T[14]}.
Our method does not generate such descriptions, as we currently only consider
single repeat units.
The above example we describe as
\texttt{155164881\_155164882insTTTTTTTGTTTTTTTTTGTTTTTTTTTTTTTTTTTTTTTTT}.
It is worth noting that all mixed repeat entries in this experiment are
effectively insertions and never contractions.

Intrigued by these dbSNP entries and to further understand these mixed repeats,
we queried the NCBI Variation Service~\cite{dbsnp2024}. Surprisingly,
$247,\!610$~entries could not be interpreted by the NCBI at all. Additionally,
$9,\!491$~entries resulted in a warning, with only a partial answer returned.
We identified that rewriting the mixed repeat into a deletion-insertion, where
the inserted sequence consists of the individual repeat units and their counts,
results in a variant that is equivalent to the input SPDI.
The example is therefore rewritten as \texttt{155164866\_155164881delins[T[23];GTTTTTTTTT[2];T[14]]}.
However, HGVS states that all repeat units need to be present in the reference
sequence. Therefore, a more conformant description is
\texttt{[155164866\_155164881T[23];\nolinebreak[0]155164881\_155164882ins[GTTTTTTTTT[2];T[14]]]}
as the trailing repeat units \texttt{GTTTTTTTTT} and \texttt{T} are not present
in the reference at this location.

\section{Discussion}\label{sec:discussion}

The HGVS descriptions produced with our method are concordant with all SNVs and
most simple insertion and deletion entries in dbSNP. For the remaining
(non-trivial) entries our method often yields insightful results.

Built on the premise that considering all LCS alignments is useful or even
required for meaningful extraction of variant descriptions, the cLCS-graph as
variant representation provides additional ways to analyze variants and could
itself be considered an alternative for textual descriptions.

Rigorous methods are required to effectively deal with ever more complex
variants both at scale and for correctness.
We argue that the focus of standardization committees, e.g., for HGVS, should be
more on comprehensibility instead of (manual) constructability.

In this section we explore some interesting examples for classes of variants as
well as limitations of our method.

\subsection{Compound Variant Descriptions}
\todo{check dat allele descriptions ergens wordt geintroduceerd}
While dbSNP is not intended for allele descriptions, we found that
there are entries described as a single deletion-insertion that consist of
multiple other smaller variants.
Often these smaller variants also exist independently in dbSNP.

Conversely, in some cases, our method results in allele descriptions that consist of many
variants, where the relatively short deletion-insertion description is arguably more
practical, e.g., \texttt{51172616\_51172630delinsACACC} vs.
\texttt{[51172616G>A;\nolinebreak[0]51172617\_51172618insA;51172620\_51172630del]}.
At the same time the value of our method is illustrated by the fact that the
substitution in our allele description is also present in dbSNP individually.
To decide a priori (or even a posteriori) whether the allele description has added
benefit is challenging, for instance
\texttt{[119251105del;119251107del;119251109del;\nolinebreak[0]119251111del;119251113del]}
vs.
\texttt{119251105\_119251113delinsAAAA}. While the deletion-insertion description is
concise, the allele description clearly communicates that this variant 
only consists of deletions. Note that when the deletion-insertion notation is preferred,
a more concise way to describe this variant is
\texttt{119251105\_119251113delinsA[4]}.
Our earlier method~\cite{vde} attempted to make the
decision of reporting a variant as a deletion-insertion when the description became too
complex. We do not do that now, because we want to make the reference and
variant structure leading instead of the simplicity of the HGVS syntax.

\subsection{Atomicity of Variants}
The individual parts of the local supremal partitioning are
independent and suggest a natural boundary for allele construction.
Determining whether a variant should be described as one or more elements has
been a long debated issue within the HGVS Variant Nomenclature Committee. A community
consultation~\cite{wg010} has been created in an attempt to resolve the discussion.
As the title ``var distance'' already suggests, it proposes ``two variants that
are separated by fewer than two intervening nucleotides (that is, not including
the variants themselves) should be described as a single ``delins'' variant.''
As there is no notion of distance between two variants, that rules out using
distance as a metric for variant separation.
When the local supremal variant consists of multiple parts, we argue
one should be hesitant to describe such variant in a different partitioning.
The independence of the parts of a local supremal variant cannot only be used for the
sake of descriptions, but also has practical implications.
The parts can be stored and processed individually and therefore allow for
optimization.

\subsection{Database Considerations}
Variants are often stored in and retrieved from databases for both clinical and
research purposes. In the case of dbSNP, this is in the form of SPDI, but in
many cases HGVS variant descriptions are used. Using HGVS variant descriptions
for retrieval has the obvious disadvantage of missing findings because of
alternative notations. The entries in dbSNP are internally represented as SPDI,
which are VOCA normalized. While their region-based
approach resembles our approach, it does not include some of the context that
is required for a complete variant representation.
\begin{figure}[ht]
\begin{center}

\begin{tikzpicture}[every node/.style={inner sep=0em, minimum height=1.4em}]
\node (a) {$\ldots$\texttt{CTA}};
\node[anchor=west, draw=orange!85, fill=orange!85] at (a.east) (b) {\texttt{CC}};
\node[anchor=west, draw=black!15, fill=black!15] at (b.east) (c) {\texttt{AAAAATAC}};
\node[anchor=west, draw=orange!85, fill=orange!85] at (c.east) (d) {\texttt{AAAAAAAAAAAAA-}\texttt{-}};
\node[anchor=west] at (d.east) (e) {\texttt{TTA}$\dots$};

\node[anchor=north] at ([yshift=-.2cm] a.south) (f) {$\ldots$\texttt{CTA}};
\node[anchor=north, draw=green!40, fill=green!40] at ([yshift=-.2cm] b.south) (g) {\texttt{CT}};
\node[anchor=north, draw=black!15, fill=black!15] at ([yshift=-.2cm] c.south) (h) {\texttt{AAAAATAC}};
\node[anchor=north, draw=green!40, fill=green!40] at ([yshift=-.2cm] d.south) (i) {\texttt{AAAAAAAAAAAAAAA}};
\node[anchor=north] at ([yshift=-.2cm] e.south) (j) {\texttt{TTA}$\dots$};

\draw ([xshift=.3em, yshift=.05cm] b.north west) -- ([xshift=.3em, yshift=.25cm] b.north west) node[anchor=south] {$\scriptstyle 12,765,037$};
\draw ([xshift=.3em, yshift=.05cm] d.north west) -- ([xshift=.3em, yshift=.25cm] d.north west) node[anchor=south] {$\scriptstyle 12,765,047$};
\draw ([xshift=.3em, yshift=.05cm] e.north west) -- ([xshift=.3em, yshift=.25cm] e.north west) node[anchor=south] {$\scriptstyle 12,765,060$};
\end{tikzpicture}%
\end{center}
\caption{
A visualization of \texttt{12765038\_12765046delinsTAAAAATACAA} from dbSNP.
Shown is a slice of the reference (top) and the observed sequence (bottom).
Deletions are highlighted in orange and insertions in green.
The supremal variant is the whole highlighted area, including the gray, where
the latter is not part of the local supremal variant
\texttt{[12765037\_12765038delinsCT;12765047\_12765059delinsAAAAAAAAAAAAAAA]}.
}
\label{fig:allele_repeat}
\end{figure}
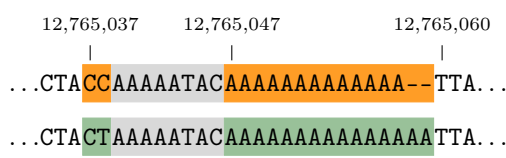

The differences are that VOCA operates on the input variant description
directly, while the input variant description for us is merely a way to
construct the observed sequence, and that VOCA is not practical for allele
descriptions. This is illustrated in Figure~\ref{fig:allele_repeat}. VOCA
identified the region~$12,\!765,\!038$--$12,\!765,\!046$ as relevant, which includes
the substitution of the~\texttt{C} to~\texttt{T} on the start position until
the end position, after which it includes the two additional~\texttt{A}s.
We on the other hand, using the supremal variant, identify the
region~$12,\!765,\!037$--$12,\!765,\!059$ as relevant. It starts one base to
the left as that~\texttt{C} can also be removed, and ends at the last of the $13$~repeating~\texttt{A}s.
The latter enables us to actually detect it as a repeat expansion.
Again, by calculating the local supremal variant, we identified two independent parts of the
variant, which are also both individually present in the database.
\subsection{Complex Repeats}

In addition to syntactic differences in repeat notation mentioned in
Section~\ref{sec:experiments}, the mixed repeat category deserves further
elaboration.
The problem of complex repeat detection in a single string has been studied before,
e.g., in~\cite{maxrep_lin_time}. The application in the context of the
description of genetic variants is still an open problem.
The fact that the NCBI Variation Service does not
accept a substantial share of the mixed repeat entries in dbSNP from the experiments also
indicates that this is unexplored territory.
As a consequence, our method currently does not result in mixed repeat
descriptions. We do, however, see value in being able to report on this type of
variant.
While the dbSNP mixed repeats clearly do not follow the HGVS recommendations,
in general the HGVS nomenclature is currently underspecified with regard to
mixed repeats.

\begin{figure}[ht]
\begin{center}
\begin{tikzpicture}[every node/.style={inner sep=0em, minimum height=1.4em}]

\node (a) {$\dots\texttt{GGA}$};
\node[anchor=west, draw=blue!40, fill=blue!40] at (a.east) (b) {\texttt{CA}};
\node[anchor=west, draw=blue!25, fill=blue!25] at (b.east) (c) {\texttt{CA}};
\node[anchor=west, draw=blue!40, fill=blue!40] at (c.east) (d) {\texttt{C}};
\node[anchor=west, draw=orange!55, fill=orange!55] at (d.east) (e) {\texttt{C}};
\node[anchor=west, draw=orange!85, fill=orange!85] at (e.east) (f) {\texttt{AC}};
\node[anchor=west, draw=orange!55, fill=orange!55] at (f.east) (g) {\texttt{AC}};
\node[anchor=west, draw=orange!85, fill=orange!85] at (g.east) (h) {\texttt{AC}};
\node[anchor=west, draw=orange!55, fill=orange!55] at (h.east) (i) {\texttt{AC}};
\node[anchor=west, draw=orange!85, fill=orange!85] at (i.east) (j) {\texttt{AC}};
\node[anchor=west, draw=orange!55, fill=orange!55] at (j.east) (k) {\texttt{AC}};
\node[anchor=west, draw=orange!85, fill=orange!85] at (k.east) (l) {\texttt{AC}};
\node[anchor=west, draw=orange!55, fill=orange!55] at (l.east) (m) {\texttt{AC}};
\node[anchor=west, draw=orange!85, fill=orange!85] at (m.east) (n) {\texttt{A}};
\node[anchor=west, draw=purple!40, fill=purple!40] at (n.east) (o) {\texttt{C}};
\node[anchor=west, draw=green!40, fill=green!40] at ([xshift=.1cm] o.east) (p) {\texttt{GC}};
\node[anchor=west, draw=green!25, fill=green!25] at (p.east) (q) {\texttt{GC}};
\node[anchor=west, draw=green!40, fill=green!40] at (q.east) (u) {\texttt{GC}};
\node[anchor=west, draw=green!25, fill=green!25] at (u.east) (v) {\texttt{GC}};
\node[anchor=west, draw=green!40, fill=green!40] at (v.east) (w) {\texttt{GC}};
\node[anchor=west, draw=green!25, fill=green!25] at (w.east) (x) {\texttt{GC}};
\node[anchor=west] at (x.east) (y) {\texttt{AAG}$\dots$};

\node[anchor=north, draw=orange!55, fill=orange!55] at ([xshift=-.3em, yshift=-.3cm] l.south) (a1) {\texttt{AC}};
\node[anchor=west, draw=orange!85, fill=orange!85] at (a1.east) (a2) {\texttt{AC}};
\node[anchor=west, draw=orange!55, fill=orange!55] at (a2.east) (a3) {\texttt{AC}};
\node[anchor=west, draw=orange!85, fill=orange!85] at (a3.east) (a4) {\texttt{A}};
\node[anchor=west, draw=purple!40, fill=purple!40] at (a4.east) (a5) {\texttt{C}};
\node[anchor=west, draw=green!25, fill=green!25] at (a5.east) (a6) {\texttt{GC}};
\node[anchor=west, draw=green!40, fill=green!40] at (a6.east) (a7) {\texttt{GC}};
\node[anchor=west, draw=green!25, fill=green!25] at (a7.east) (a8) {\texttt{GC}};

\draw
    ([xshift=.3em, yshift=.05cm] b.north west) -- ([xshift=.3em, yshift=.25cm] b.north west) node[anchor=south, xshift=-.2cm] {$\scriptstyle 63,493,066$}
    ([xshift=.3em, yshift=.05cm] f.north west) -- ([xshift=.3em, yshift=.25cm] f.north west) node[anchor=south, xshift=.2cm] {$\scriptstyle 63,493,072$}
    ([xshift=.3em, yshift=.05cm] p.north west) -- ([xshift=.3em, yshift=.25cm] p.north west) node[anchor=south] {$\scriptstyle 63,493,090$}
    ([xshift=.3em, yshift=.05cm] y.north west) -- ([xshift=.3em, yshift=.25cm] y.north west) node[anchor=south] {$\scriptstyle 63,493,102$}
    (o.south east) -- (a1.north west)
    (p.south west) -- (a8.north east)
;
\end{tikzpicture}%
\end{center}
\caption{
A visualization of the mixed repeat \texttt{63493089CA[4]CG[3]C[1]} on a slice of
the reference sequence, effectively describing an insertion of $4\times\texttt{AC}$ highlighted in
orange and
$3\times\texttt{GC}$ highlighted in green described in HGVS as \texttt{63493089\_63493090ins[AC[4];GC[3]]}.
Both the \texttt{AC} and \texttt{GC} also occur in the reference. Note that
these patterns partially overlap as highlighted in purple.
The supremal variant identifies an additional adjacent relevant occurrence of an unchanged
\texttt{CA} repeat highlighted in blue.
}
\label{fig:mixed_repeat}
\end{figure}
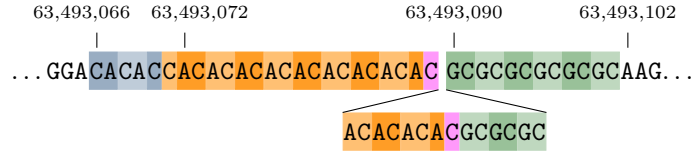

\noindent
We turn to Figure~\ref{fig:mixed_repeat} for an example to guide the mixed
repeats discussion.
The dbSNP HGVS description \texttt{63493089CA[4]CG[3]C[1]} has a corresponding SPDI
entry \texttt{63493088:C:CACACACACGCGCGC}. This entry is effectively described as an
insertion, and the VOCA region does not span the existing repeat units in the reference.
It therefore begs the question if the reference actually plays a role here, or
if the repeat notation is based on the inserted sequence only.
The supremal variant includes all the repeat units and the
adjacent \texttt{CA} repeat starting from $63,\!493,\!066$.
The description can start either at $63,\!493,\!071$ with the \texttt{CA} and \texttt{CG}
rotation or at $63,\!493,\!072$ with the \texttt{AC} and \texttt{GC} rotation.
With that context, we can expand both the \texttt{AC}/\texttt{CA} and
\texttt{CG}/\texttt{GC} repeat units to $13$ and $9$, respectively.
HGVS dictates the latter because of its shifting rule:
\texttt{63493072\_63493089AC[13]} and \texttt{63493090\_63493101GC[9]}.
Combining the two into the current mixed repeat format results in
\texttt{63493072\_63493101AC[13]GC[9]}, conveniently getting rid of the
\texttt{C[1]} from the original entry as there is no such repeat in the
reference. The disadvantage of this format is that
we cannot determine how many repeat units there were originally in the reference.
An alternative way to address that issue is to describe this variant as allele
\texttt{[63493072\_63493089AC[13];63493090\_63493101GC[9]]}.
Note that patterns \texttt{AC} and \texttt{CG} partially overlap, which in
general leads to multiple equivalent descriptions.

\subsection{Biological Features}
With our current methods, inversions are not detected natively. However, when
the shortest path to describe an inversion is a single replacement, we attempt
to see if that replacement can also be expressed as an inversion. For the
concrete case of our experiment on dbSNP, we are able to trivially identify
all inversions when applying this strategy to the supremal variant. Unfortunately,
when an inversion is more aptly represented as an allele description it cannot
be identified without additional efforts. Our former extraction method~\cite{vde}
performs an alignment against both the forward and reverse complement of a
reference sequence. A similar technique might be applied to the alignment
described in Section~\ref{sec:methods}.

So far we have reported only on chromosomal reference sequences, yielding genomic
variant descriptions in HGVS. In practice, variants in HGVS are often
reported relative to coding transcript references. While the variant descriptions
resulting from our methods can be mapped to transcript references (using, for
instance, the Mutalyzer tool suite~\cite{mutalyzer}), our extraction methods do
not take the features, i.e., introns and exons, of these transcripts into account.
This implies that for variants on the exon-intron boundary, equivalent
variant descriptions on genomic level could have different effects on the
transcript.

\subsection{Performance Challenges}
The nature of the alignment method can lead to (sometimes counterintuitively)
large graphs. This presents performance challenges for larger variants, where
memory usage grows significantly. This is especially true for large
deletions, which have a large number of paths, while having compact
descriptions.
While the current Python implementation is adequate to perform a single
experiment on chromosomal scale as in Section~\ref{sec:experiments}, for
routine whole genome sequencing analysis this implementation could be
considered too slow. For these situations a reimplementation in a more
performance-oriented language might be in order.

\todo{list largest graph?}

\section{Conclusions}\label{sec:conclusion}

In this paper we addressed the problem of finding suitable representations for
genetic variants and subsequently choosing particular descriptions for each
variant.
Beyond the crucial property that variant descriptions need to be unambiguously
interpretable, there is a clear need for directions on how to describe them in
various scenarios.
The introduced cLCS-graph data structure is a complete representation of a
variant. It enables the three extraction methods presented in this paper that
address the needs for usage of variant descriptions ranging from clinical
reporting to databases. Experiments clearly show the benefits of our methods.

Together with the comparative capabilities of the variant algebra, one is
ultimately free in how to choose a variant description while still being able
to relate them to other notations of the same variant or other variants
altogether.
Although there is not always clear added benefit for writing variants as
alleles, our method does enable this through local supremal variant extraction.
Besides practical value as demonstrated in the experiments, our deterministic
extraction methods also support rational choices for Variant Nomenclature
Standard processes.

\subsection{Future Work}

Our methods are currently agnostic to genomic features and could be enriched to
enable exon-intron-aware variant extraction. Inversions and transpositions could be enabled
analogous to the way inversions are dealt with in~\cite{vde}. Earlier experiments showed that there
are many variants in dbSNP that have a non-disjoint relation with one or more
other variants in the database. By extending dbSNP with our concepts it would
enable richer query possibilities and allele descriptions.

\todo{variant description in context of other variant (from redmar)}

\section{Code Availability}
A Python implementation is available at
\url{https://github.com/mutalyzer/algebra/tree/v1.5.2} with a snapshot
deposited at \url{https://doi.org/10.5281/zenodo.17241368} and integrated in a
web interface at \url{https://test.mutalyzer.nl/normalizer}.

\printbibliography

\end{document}